\documentclass{jfm}

\usepackage{graphicx}
\usepackage{newtxtext}
\usepackage{float}
\usepackage{newtxmath}
\usepackage{booktabs}
\usepackage{amsmath}
\usepackage{natbib}
\usepackage{adjustbox}
\usepackage{hyperref}
\usepackage{changes}  
\usepackage{physics}
\usepackage{derivative}
\usepackage{pdfpages}
\usepackage{pgfplots}
\pgfplotsset{compat=newest}
\usepackage{lscape} 
\usepackage{adjustbox} 
\usepackage{caption}
\usepackage{subcaption}
\usepackage{graphicx}
\usepackage{comment}
\usepackage{subcaption}
\usepackage{multirow}
\usepackage{lscape}
\usepackage{overpic}
\usepackage{mathtools}
\usepackage{float}
\usepackage{placeins}
\hypersetup{
    colorlinks = true,
    urlcolor   = blue,
    citecolor  = black,
}

\newcommand{\RomanNumeralCaps}[1]
\linenumbers
\usepackage[justification=raggedright,singlelinecheck=false,width=\textwidth]{caption}
\usepackage{tikz}

\newcommand\St{\mbox{\textit{St}}}  
\newcommand\Nu{\mbox{\textit{Nu}}} 

\title{Modelling friction and heat transfer in turbulent forced convection over porous lattices}

\author{Aneek Chakraborty\aff{1}
  \corresp{\email{a.chakraborty@tudelft.nl}},
  Stefan Hickel\aff{1}
 \and Davide Modesti\aff{2}}

\affiliation{
    \aff{1}Aerodynamics Group, Faculty of Aerospace Engineering, Delft University of Technology, Kluyverweg 2, 2629 HS Delft, The Netherlands,
    \aff{2}Gran Sasso Science Institute, viale Francesco Crispi 7, L'Aquila, Italy
}
\begin{document}
\maketitle

\begin{abstract}
We perform direct numerical simulations (DNS) to investigate how cubic-lattice porous substrates influence momentum and heat transfer in turbulent channel flows. The simulations span friction Reynolds numbers from 260 to 1500, Prandtl numbers of 0.5, 1, and 2, and substrate porosities of 50\%, 71\%, and 87\%. We show that theories developed for rough-wall turbulence can be extended to porous surfaces by replacing the roughness height with the inverse of the streamwise Forchheimer coefficient. The shifts in the mean velocity and temperature profiles follow existing fully rough momentum and thermal theories, enabling their prediction with rough-wall models. Combining these models with synthetic temperature and velocity profiles, we derive analytical formulas for the friction coefficient and Stanton number that agree with our DNS data to within 5\%. The performance enhancement factor, which measures heat-transfer augmentation relative to the pressure-drop penalty at constant pumping power, is comparable to that obtained for rough surfaces. This suggests that porous substrates provide an alternative method for enhancing heat transfer in turbulent flows.
\end{abstract}

\begin{keywords}
Authors should not enter keywords on the manuscript, as these must be chosen by the author during the online submission process and will then be added during the typesetting process (see \href{https://www.cambridge.org/core/journals/journal-of-fluid-mechanics/information/list-of-keywords}{Keyword PDF} for the full list).  Other classifications will be added at the same time.
\end{keywords}

{\bf MSC Codes }  {\it(Optional)} Please enter your MSC Codes here

\section{Introduction}

Turbulent forced convection in internal flows is fundamental to several industrial applications, such as fuel cells, nuclear plants, turbine blades, and rocket nozzles, where heat exchangers play a pivotal role. In all these cases, heat exchangers must maximize heat transfer. Towards this end, surface texturing enhances turbulent mixing and has become a widely accepted practice in industry. However, this comes at the cost of a higher pressure drop~\citep{Ligrani}. Increasing heat transfer with a marginal or no pressure-drop penalty is challenging because roughness has a major pressure-drag contribution, which has no equivalent in heat transport. Riblets are perhaps the only surface pattern that can invert this trend, although the resulting increase in heat transfer remains limited~\citep{Rouhi_Endrikat_Modesti_Sandberg_Oda_Tanimoto_Hutchins_Chung_2022}.

From a fundamental perspective, we have reached a consensus on several aspects of turbulent heat transfer over rough surfaces, including the heat-transfer increase caused by roughness~\citep{Owen_Thomson_1963, Dipprey1963, Brutsaert1975,    MacDonald_Hutchins_Chung_2019, PEETERS2019454, Chungreview}. This can be measured using the viscous-scaled downward shift of the mean temperature profile, expressed by the temperature-deficit term relative to the smooth wall, $\Delta \Theta^+=\Theta_S^+-\Theta_R^+$, which plays the same role for heat transfer as the Hama roughness function $\Delta U^+=U_S^+-U_R^+$ does for momentum,
\begin{subequations}
\begin{align}
U_R^+ &= \frac{1}{\kappa}\log{y^+} + A - \Delta U^+ (\ell^+), \label{eq:vel_R}\\
\Theta_R^+ &= \frac{1}{\kappa_\theta}\log{y^+} + A_\theta(\Pran) - \Delta \Theta^+ (\ell^+,\Pran),\label{eq:temp_R},
\end{align}
\end{subequations}
where $\kappa=0.387$ and $A=4.17$ are the von K\'arm\'an constant and the additive constant for the mean velocity in channel flow. For temperature, $\kappa_\theta=0.459$, and the smooth-wall intercept $A_\theta(\Pran)$ depends on the Prandtl number $\Pran=\nu/\alpha$,
\begin{equation}\label{calcBeta}
     A_{\theta}(Pr) = \frac{1}{\kappa_{\theta}} \left [\frac{2\pi\ C_{\theta}^{2/3}}{3\sqrt{3}} Pr^{2/3} + \frac{1}{3} \mathrm{log} Pr - \left(\frac{1}{6} +\frac{1}{2\sqrt{3}} +\frac{2}{3} \mathrm{log}\ C_{\theta} -\mathrm{log}\ k_{\theta}  \right)\right] ,
\end{equation}
where $C_{\theta} = 10$~\citep{pirozzoli_23,Pirozolli_Modesti_2024}, and $\nu$ and $\alpha$ are the fluid viscosity and thermal diffusivity. Here, $\Theta=T-T_w$ is the temperature difference, where $T_w$ is the wall temperature. The plus superscript indicates quantities normalized in wall units, namely using the friction velocity $u_\tau=\sqrt{\tau_w/\rho}$, the viscous length scale $\delta_v=\nu/u_\tau$, and the friction temperature $\theta_\tau=q_w/u_\tau$, where $\rho$ is the fluid density, $\tau_w$ is the wall shear stress, and $q_w$ is the wall heat flux. We use capital letters to indicate Reynolds averages in time and homogeneous spatial directions, and lowercase letters for the corresponding fluctuations. Subscripts S and R denote the smooth and rough/porous walls, respectively.

The velocity and temperature shifts depend on the geometry of the rough surface or porous layer and on the viscous-scaled Reynolds number $\ell^+$, where $\ell$ is a suitable length scale of the surface texture, typically the roughness height for rough surfaces. The temperature shift further depends on the Prandtl number~\citep{Chungreview}. Equation \eqref{eq:Brutsaert} presents an alternative formulation using the interfacial temperature $\Theta_i^+$, which is the temperature at the wall-normal height $y_i$ where the roughness sublayer ends and the rough-wall temperature becomes logarithmic, given by:
\begin{equation}\label{eq:Brutsaert}
    \Theta_R^+ = \frac{1}{\kappa_{\theta}}\mathrm{log}(y/y_i) + \Theta_i^+(\ell^+,Pr).
\end{equation}

For rough surfaces, \citet{Brutsaert1975} modelled the interfacial temperature as $\Theta_i^+ \sim (k^+)^{1/4}Pr^{1/2}$, a scaling that has recently been confirmed by \citet{Zhong_Hutchins_Chung_2023} using direct numerical simulation (DNS) data over sinusoidal roughness from \citet{MacDonald_Hutchins_Chung_2019} and irregular roughness from \citet{PEETERS2019454}.

From this model, it follows that the temperature shift $\Delta\Theta^+$ increases with $\ell^+$ up to a maximum, beyond which it decreases, with a consequent reduction in global heat transfer. This model has been confirmed by DNS results, at least for some values of the Prandtl number and some surface types~\citep{Zhong_Hutchins_Chung_2023}.

While heat transfer over rough surfaces has been studied by several authors~\citep{Forooghi2018,Fugmann2019,Yang2025}, less is known about permeable surfaces, and an open question is which length scale should be used to characterise the substrate.

In the case of rough surfaces, the typical length scale is the roughness height; however, for porous surfaces, different options are possible. Besides obvious geometrical length scales, such as the mean pore size, other possibilities considered in the literature stem directly from the generalised Darcy--Forchheimer law,

\begin{equation}
-\frac{\partial p}{\partial x_j} = \left[\mu K_{ij}^{-1} + \rho\left(u_ku_k\right)^{1/2}\alpha_{ij}\right]u_i,
\end{equation}

where $\partial p/\partial x_j$ is the pressure drop through the porous medium in the $j$-th direction, $\mu$ and $\rho$ are the dynamic viscosity and density of the fluid, $u_i$ is the velocity component in the $i$-th direction, $K_{ij}$ is the Darcy permeability tensor, and $\alpha_{ij}$ is the Forchheimer permeability tensor. Most previous studies consider only the Darcy contribution to the pressure drop, under the assumption that the pore size is small and inertial effects are negligible. For this reason, a common choice is to use the square root of the Darcy permeability in the dominant flow direction, $\sqrt{K_{ii}}$, as a reference length scale. This approach was used by \citet{Breugem} and \citet{breugem_boersma_uittenbogaard_2006}, who conducted DNS of turbulent flows over a porous layer modelled using volume-averaged equations, and by \citet{Habibi_Khorasani_Luhar_Bagheri_2024}, who conducted DNS of turbulent flows over cubic porous lattices.

Experimental studies on porous surfaces with direct drag measurements are available for different types of porous surfaces~\citep{SUGA2010974, Manes_Poggi_Ridolfi_2011, SUGA2011586, Suga_Nakagawa_Kaneda_2017,esteban_22,wangsawijaya_23,vijay_24,hartog_24}, and the widely used approach is to use $\sqrt{K_{ii}}^+$ as a reference length scale. Some authors pointed out that the grazing flow is modified by both the surface roughness of the uppermost layer and the permeability characteristics of the porous substrate~\citep{wangsawijaya_23,esteban_22}. They proposed a length scale that combines the `roughness' and `permeability' components. While these studies have advanced the characterisation of momentum transfer over porous surfaces, the corresponding thermal behaviour remains comparatively less explored. Experiments on heat transfer over rough surfaces have shown that accurately characterizing thermal structures at the wall demands careful measurement techniques~\citep{AbuRowin2024}, particularly when resolving thermal fluctuations in a turbulent boundary layer~\citep{Foroozan2025}. Such experiments exploring heat transfer become more challenging for porous surfaces, where fluid penetration introduces sub-surface transport that is hard to access with conventional wall-based techniques.

In some recent studies by our group~\citep{Shahzad2022,Shahzad_Hickel_Modesti_2023,shahzad_25}, we performed DNS of turbulent flows grazing over perforated plates and found that inertial effects inside the pores are significant, with the inverse of the wall-normal Forchheimer coefficient, $1/\alpha_{yy}^+$, emerging as the dominant length scale.
Numerical studies using modelled boundary conditions for the porous substrate have in some cases reported drag reduction compared to the smooth-wall reference~\citep{Rosti_Brandt_Pinelli_2018,Gomez-de-Segura2018}; however, all recent pore-resolved numerical simulations~\citep{Kuwata_Suga_2017, Shahzad_Hickel_Modesti_2023, Habibi_Khorasani_Luhar_Bagheri_2024, Hao_Garcia-Mayoral_2025} and experiments with direct drag measurements~\citep{wangsawijaya_23,vijay_24,hartog_24} report increased drag for porous surfaces compared to a smooth wall. At sufficiently high pore Reynolds number, the emergence of a fully rough regime has been consistently observed~\citep{wangsawijaya_23,Shahzad_Hickel_Modesti_2023}; however, the determination of a dominant flow length scale is still debated.

While the influence of permeable walls on the overlying momentum field has been extensively studied, their effect on heat transfer remains less explored. Most studies in the literature analysing convection in porous channels have used the macroscopic Darcy-Brinkmann-Forchheimer equation to model the flow in the porous medium \citep{Vafai_1984,VAFAI19871391, LEE1999423,ALKAM2001931, YANG20092956, HUANG20101164,AGUILARMADERA20111355, VAFAI1990254}. Pore-resolved simulations including heat transfer are scarce in the literature.

\citet{Chandesris2013} performed DNS over an array of cubes representing a porous wall layer and showed that wall permeability promotes Kelvin–Helmholtz-type large-scale structures that enhance heat transfer across the porous interface. \citet{Motoki_Tsugawa_Shimizu_Kawahara_2022} explored how similar large-scale spanwise structures drives heat transfer in a permeable channel flow to the diffusivity-independent 'ultimate regime'. More recently, \citet{NishiyamaKuwataSuga} carried out DNS of turbulent flow over a simple cubic structure with conjugate heat transfer, conducting a parametric study in which the directional permeabilities were progressively varied to approach a nominally isotropic medium. They found that treating the porous substrate as isothermal yields weak temperature mixing within the porous geometry, underscoring the importance of thermal coupling between the fluid and solid phases. In related work, \citet{Brethouwer_Habibi_Khorasani_Bagheri_2026} performed conjugate heat transfer simulations over a porous cubic lattice and reported an unfavorable departure from the Reynolds analogy at higher volume porosities. 

The available literature provides several models for predicting the additional drag induced by porous and rough surfaces~\citep{Yang_Sadique_Mittal_Meneveau_2016,Meneveau_Hutchins_Chung_2024, Abu_Rowin_Zhong_Saurav_Jelly_Hutchins_Chung_2024}, whereas analogous models for heat transfer remain lacking. In this work, we use pore-resolved direct numerical simulations (DNS) to investigate momentum and heat transfer over porous substrates and extend predictive models for the heat transfer coefficient, originally developed for rough surfaces, to porous media. To this end, we develop a comprehensive database of turbulent channel flows over simple-cubic porous substrates, spanning friction Reynolds numbers $\Rey_\tau\approx250$–$1500$, volume porosities $\Phi=0.5,0.71,0.87$, and Prandtl numbers $\Pran=0.5,1,2$.

\section{Methodology}\label{Methodology}

\begin{figure}
    \centering
    \begin{overpic}[width=\textwidth]{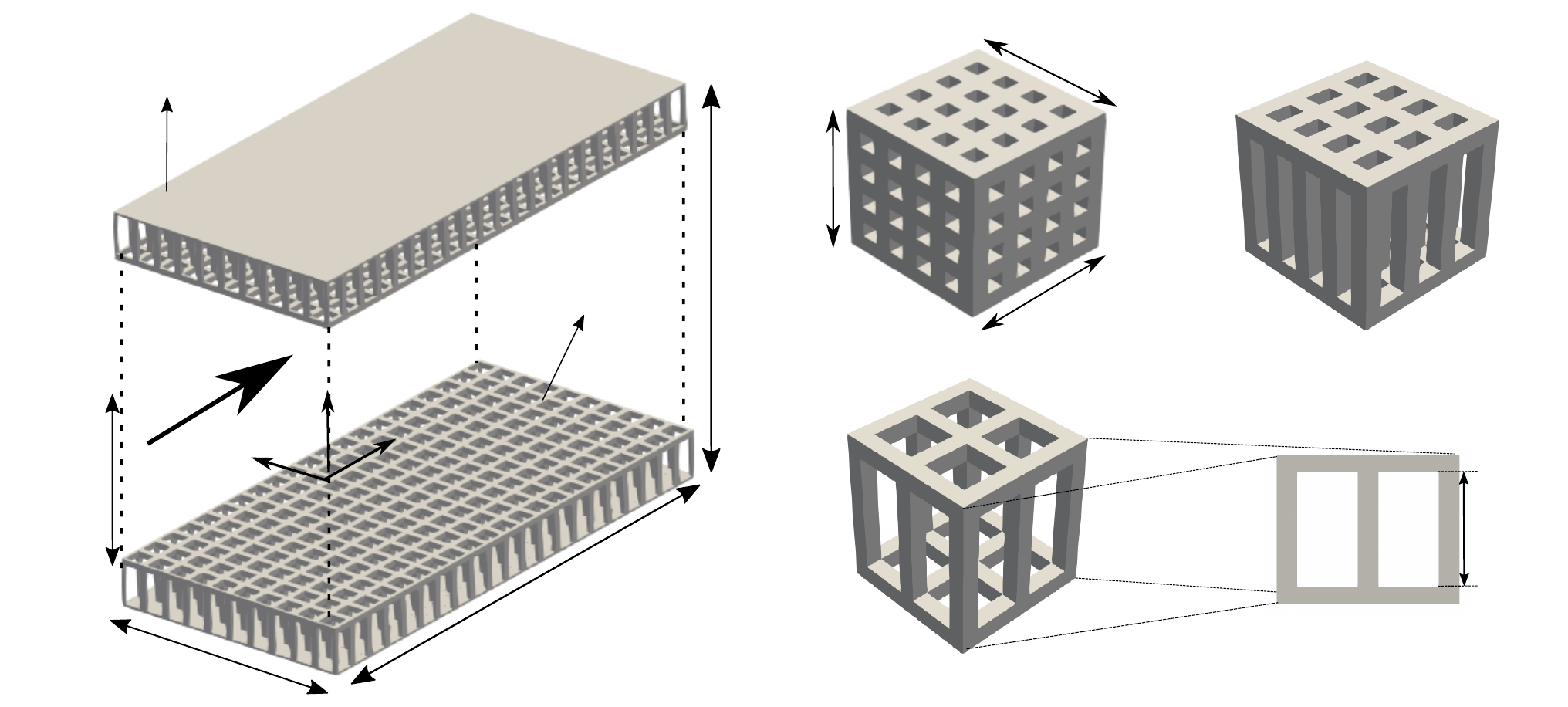}
     \put(4,45){$(a)$}
     \put(12,1){\small $L_z$}
        \put(33,5.5){\small $L_x$}
        \put(46,26){\small $L_y$}
        \put(8,21){\textbf{Flow}}
        \put(33,28.5){\small porous}
        \put(33,26.5){\small substrate}
        \put(4,13){ $\delta$}
        \put(4,6){ \small $h$}
        \put(26,16) {$\mathbf{x}$}
        \put(19,20) {$\mathbf{y}$}
        \put(13,14.5) {$\mathbf{z}$}
        \put(8.5,39){\small wall}
     \put(48,45){$(b)$}
        \put(50,40){\small (i)}
        \put(76,40){\small (ii)}
        \put(50,20){\small (iii)}
        \put(48,33){\small $0.3\delta$}
        \put(68,41){\small $0.3\delta$}
        \put(67,24){\small $0.3\delta$}
        \put(94,11){\small $d_i$}
    \end{overpic}
    \caption{\textit{(a)} Schematic of the computational domain with dimensions $L_x\times L_y\times L_z = 3\delta\times 2.6\delta\times 1.5\delta$). \textit{(b)} Unit cells of porous geometries used for simulations for (i) $50\%$, (ii) $71\%$, and (iii) $87\%$. The unit cell has dimension $0.3\delta\times 0.3\delta\times 0.3\delta$ and $d_i$ is the pore size in the i-th direction.}
    \label{fig:SCCgeo}
\end{figure}
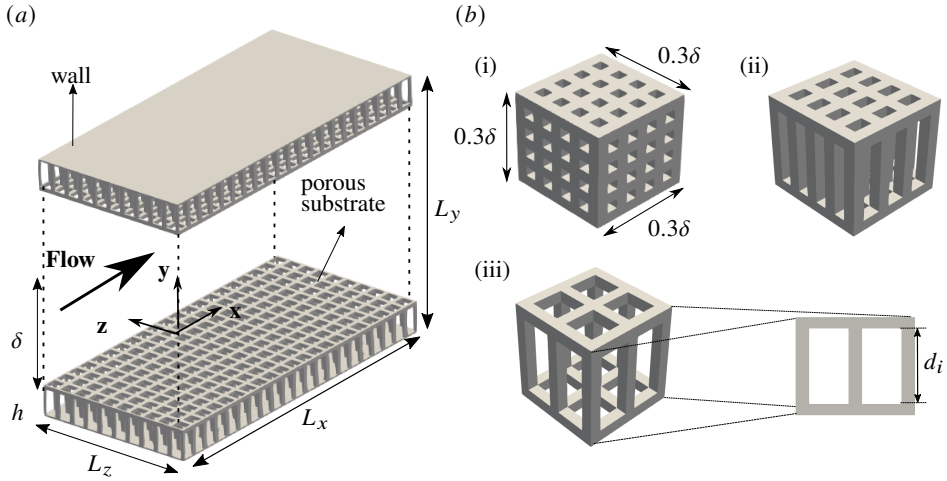

We solve the incompressible Navier--Stokes equations, augmented with three equations for passive temperature transport, using a second-order finite-difference method on a staggered mesh, which guarantees exact discrete conservation of kinetic energy and temperature variance in the inviscid limit, in addition to conserving momentum~\citep{orlandi_00}. The Poisson equation for pressure is solved in spectral space in the homogeneous directions, whereas a second-order finite-difference discretisation is used in the wall-normal direction. For time integration, a third-order, three-stage Runge--Kutta algorithm by Alan Wray is employed. The temperature is modelled as a passive scalar field, and for each flow case we consider three values of the Prandtl number, $\Pran=0.5,1,2$. The streamwise momentum equation is forced to maintain a constant mass flow rate. Similarly, a uniform volumetric heat flux is used to maintain a constant bulk temperature. Statistics are collected for at least $\Delta T u_\tau/h=50$. Furthermore, statistics are intrinsically averaged, i.e. the averages within the porous medium consider only the fluid domain.

Cubic-lattice structures are used as porous substrates, as shown in figure \ref{fig:SCCgeo}, following the work of \citet{Kuwata_Suga_2017} and \citet{Habibi_Khorasani_Luhar_Bagheri_2024}. The ligaments have a square cross-section with side length $d_{l}= 0.039\delta$, which is kept fixed throughout this study. The pore sizes, denoted by $d_{x}$, $d_{y}$, and $d_{z}$ in the streamwise, wall-normal, and spanwise directions, respectively, are systematically varied to alter the volume porosity $\Phi$, defined as the open-to-solid volume ratio. 

Three volume porosities are considered in the present work: $50\%$, denoted by $L$; $71\%$, denoted by $M$; and $87\%$, denoted by $H$. For each porosity, we consider three friction Reynolds numbers, except for the high-porosity case, for which we study four Reynolds numbers, up to $\Rey_\tau\approx1500$. 
We also performed baseline smooth-wall simulations at approximately matching Reynolds numbers, as reported in table~\ref{tab:simulation_campaign_vertical}. The streamwise and spanwise directions are treated with periodic boundary conditions, whereas no-slip, isothermal boundary conditions are imposed at the solid walls.

The solid porous medium is represented with an immersed boundary method, in which the viscous and conductive derivatives at the first point off the wall are calculated using the wall distance instead of the local mesh spacing. Points that fall too close to the wall are corrected using a point-implicit time-stepping scheme, as described in the recent work by~\citet{luchini_25}. The same algorithm has been used to study turbulent pipe flow over rough surfaces~\citep{De_Maio_Latini_Nasuti_Pirozzoli_2023}.

A uniform grid spacing is used in the periodic directions, whereas the grid-stretching function proposed by \citet{PIROZZOLI2021110408} is used to cluster the points in the wall-normal direction near the interface with the porous medium. A geometric progression is used to increase the mesh spacing towards the bottom of the porous substrates. The mesh spacing in the wall-parallel directions guarantees approximately seven grid points per ligament while satisfying the restriction imposed by the Batchelor scale for the temperature field at $\Pran=2$. This mesh spacing was selected following a mesh-refinement study for the present type of porous surface, see Appendix~\ref{app:gridConvergence}.

Simulations are performed in a domain with the dimensions $3\delta\times 2(\delta+h)\times 1.5\delta$, except for cases $S1500$ and $H1500$, which were performed in a larger domain $6\delta\times 2(\delta+h)\times 3\delta$, where $\delta$ is the half-channel height and $h$ is the height of the porous substrate, as shown in figure \ref{fig:SCCgeo}. This domain size is smaller than that typically used for smooth-wall simulations, but larger than that typically used for rough-wall simulations~\citep{MacDonald_Hutchins_Chung_2019, Zhong_Hutchins_Chung_2023}, and it has been used to simulate other types of porous surfaces~\citep{Shahzad_Hickel_Modesti_2023} and roughness~\citep{digiorgio_20}.
Here, the overbar symbol represents Reynolds averaging in the homogeneous directions and in time for fluctuation statistics.

To characterise the porous lattices, we additionally carried out Stokes-flow simulations using the OpenFOAM solver, whose setup is reported in Appendix~\ref{app:Openfoam}. The Darcy and Forchheimer permeability values were calculated from the pressure drop, and their viscous-scaled values are reported in table \ref{tab:simulation_campaign_vertical}.
\begin{landscape} 
\begin{table}
\centering
\renewcommand{\arraystretch}{2}
\adjustbox{max height=0.5\textheight, max width=1.5\textwidth}{
\begin{tabular}{ccccccccccccccccccc}
\toprule
  Cases & $Re_{b}$ & $Re_{\tau}$ & $\Delta x^+$ & $\Delta z^+$ & $\Phi$ & $d_x^+$ & $d_y^+$ & $d_z^+$ & $\sqrt{K_x^+}$ & $\sqrt{K_y^+}$ & $\sqrt{K_z^+}$ & $1/\alpha_x^+$ & $1/\alpha_y^+$ & $1/\alpha_z^+$ &$C_f \times 10^2$ & $Nu (Pr=2)$ & $Nu (Pr=1)$ & $Nu (Pr=0.5)$\\

S260 &8800&262.08&  6.14 & 3.07 & -&-  & -& -& -& -& -& -&-& -&0.69&42.60&31.52&23.02\\
S500 &18800&509.27& 7.96 & 3.98 &- &- &- &- &- &- &- &- &-& -&0.58&80.61&57.69&40.27\\
S700 &26000&676.34&  9.38 & 4.69 &- &- &- &- &- &- &- &- &- &-&0.54&105.39&74.43&51.05\\
S1500 &63000&1497.86&  4.69 & 4.69 &- &- &- &- &- &- &- &- &- &-&0.45&216.77&149.93&98.97\\
L260 &9100&275.82&  1.48 & 1.48 & $50\%$ & 10.73 & 9.96 & 10.73 & 1.04 & 1.12 & 1.04 & 1.11 & 0.79 & 1.11  &1.04&55.21&40.02&28.84\\
L500 &18200&511.91&  3.03 & 3.03 & $50\%$ & 18.77 & 19.91 & 18.49 & 1.91 & 2.06 & 1.91 & 2.06 & 1.46 & 2.06 &0.90&105.03&71.55&48.56\\
L700 &24960&687.64&  4.24 & 4.25 & $50\%$ & 26.75  &24.31 &26.75 &2.59 &2.79 &2.59 &2.77 &1.97 &2.77 &0.86&142.57&94.66&62.68  \\
M260 &7800&258.06&  1.43 & 1.47 & $71\%$ &  15.77 & 67.38 & 10.03 & 1.78 & 2.32 & 3.09 & 2.63 & 3.14 & 4.78  &1.35&62.51&42.64&29.47\\
M500 &14820&500.69&  2.75 & 2.8 & $71\%$ & 30.59 & 130.73 & 19.47 & 3.44 & 4.51 & 6.01 & 5.11 & 6.09 & 9.28  &1.39&143.26&88.18&55.39\\
M700 &21060&690.88& 2.02 & 2.06 & $71\%$ & 42.22 &180.59 &26.86 &4.77 &6.21 &8.27 &7.05 &8.41 &12.79 &1.31&200.24&120.50&73.61  \\
H260 &4940&245.65& 1.49 & 1.49 & $87\%$ &  27.29 & 64.14 & 27.29 & 5.31 & 4.53 & 5.31 & 14.13 & 8.43 & 14.13 &3.24&110.84&67.22&40.66\\
H500 &10140&506.19&  2.87 & 2.87 & $87\%$ &  56.24 & 132.17 & 56.24 & 11.08 & 9.45 & 11.08 & 29.13 & 17.56 & 29.13 &3.29&235.02&144.94&62.68\\
H700 &13988& 694.09&  2.14 & 2.03 & $87\%$ & 77.10 & 181.22 &77.10 &15.00 &12.79 &15.00 &39.94 &23.85 &39.94 &3.29&318.68&201.00&117.28 \\
H1500 &29380&1501.41&  4.09 & 3.89 & $87\%$ & 166.273 & 390.79&166.273 &32.35 &27.58 &32.35 &86.12 &51.36 &86.12 &3.39&592.63&380.75&227.63 \\
\bottomrule
\end{tabular}}
\caption{Direct numerical simulation dataset, including smooth and porous wall cases. The volume porosity is indicated by $\Phi$, and $\Rey_b = 2(\delta+h) U_b/\nu$ is the bulk Reynolds number. The viscous-scaled grid spacings in the wall-parallel directions are denoted by $\Delta x^+$ and $\Delta z^+$ and pore spacings are represented by $d_x$, $d_y$, and $d_z$. The viscous-scaled Darcy permeabilities in the streamwise, wall-normal and spanwise direction are given by $\sqrt{K_x^+}$, $\sqrt{K_y^+}$ and $\sqrt{K_z^+}$, whereas $\alpha_x^+$,  $\alpha_y^+$, and  $\alpha_z^+$ are the viscous-scaled Forchheimer permeabilities. The friction coefficient  $C_f = 2/(U_b^+)^2$ and the Nusselt number $Nu = \St \Rey_b \Pran $ for all $\Pran$ case are also reported.}.
\label{tab:simulation_campaign_vertical}
\end{table}
\end{landscape}

\section{Results}\label{Results}

\subsection{Instantaneous flow field}\label{FlowVisu}

\begin{figure}
    \centering
    \includegraphics[width = \textwidth]{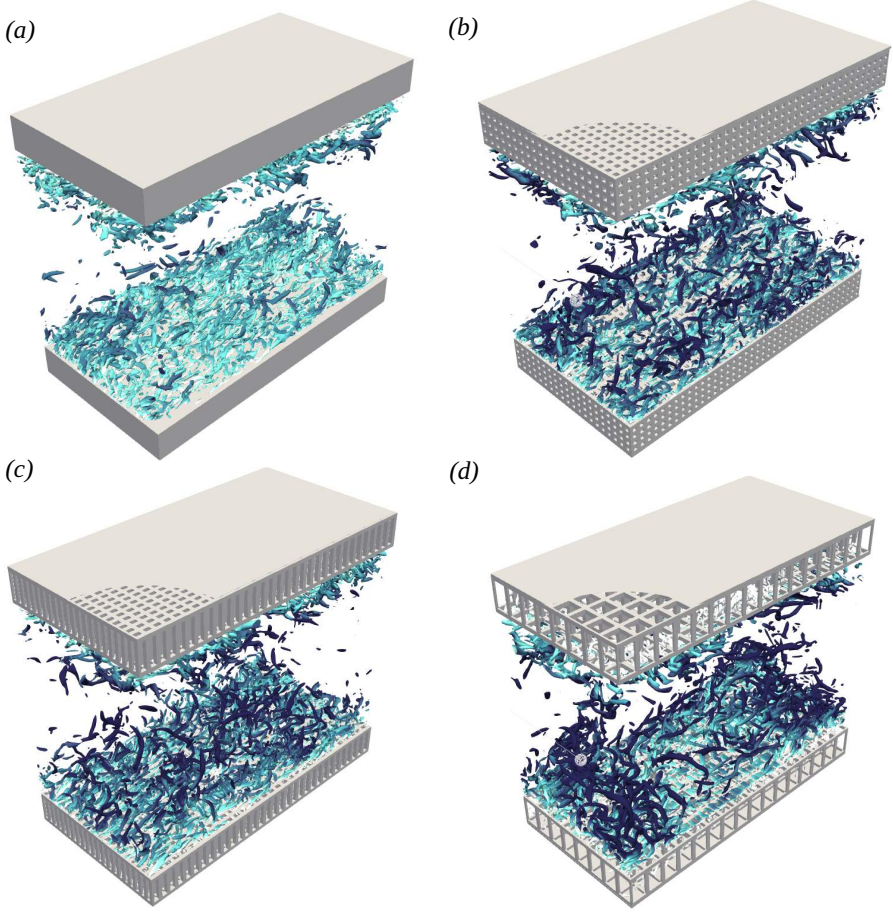}
    \caption{Isometric view of instantaneous flow field for $S700$ \textit{(a)}, $L700$ \textit{(b)}, $M700$ \textit{(c)} and $H700$\textit{(d)}, visualized using Q-criterion colored by the streamwise velocity.}
    \label{fig:combinedplots}
\end{figure}

We begin by examining the three-dimensional instantaneous flow field in figure \ref{fig:combinedplots}, which shows isosurfaces of the Q-criterion at $Re_{\tau}\approx 700$ for different porous substrates and the smooth-wall case. Over the smooth wall, the vortical structures are predominantly elongated in the streamwise direction and inclined with respect to the wall, consistent with the action of the mean shear. Increasing the volume porosity leads to a visibly different vortex organization. The vortices become less elongated and exhibit a weaker preferential orientation, indicating a substantial departure of the near-wall flow organization from that of the smooth-wall case.

\begin{figure}
    \centering
    \includegraphics[width=\linewidth]{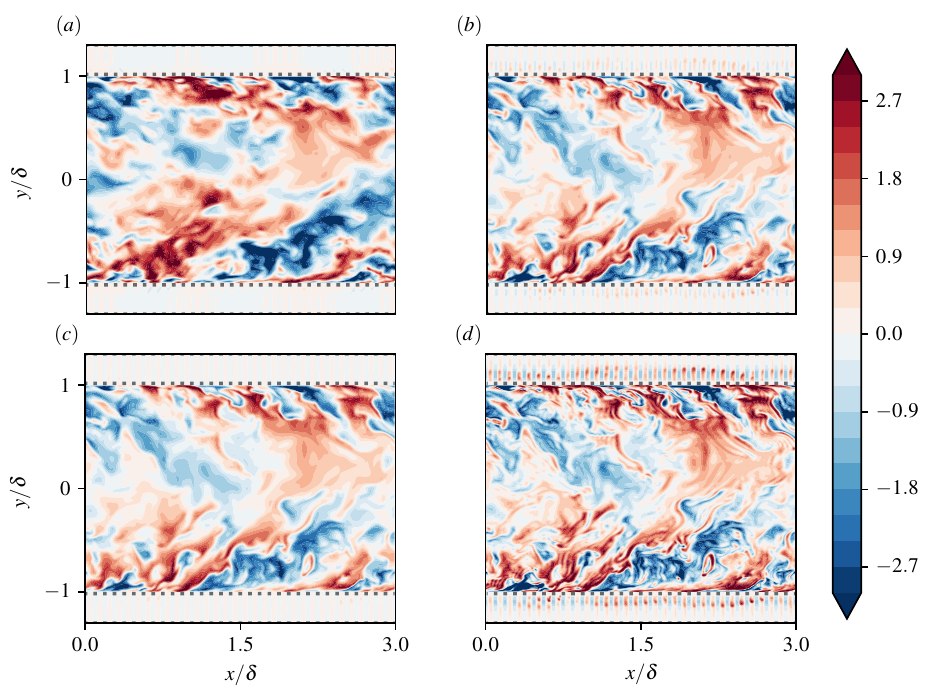}
    \caption{Instantaneous streamwise velocity fluctuations $u/u_{\tau}$ \textit{(a)}, temperature fluctuations $\theta/\theta_{\tau}$ at $Pr=1$ \textit{(b)}, $Pr=0.5$ $\textit{(c)}$ and $Pr=2$ $\textit{(d)}$ in a longitudinal plane for case $M700$.}
    \label{fig:udashthetadash_xy}
\end{figure}

Figure \ref{fig:udashthetadash_xy} shows instantaneous contours of the streamwise velocity and thermal fluctuations at different Prandtl numbers for flow case M700. At $\Pran=1$, we observe the presence of large-scale temperature and velocity structures spanning the entire channel from the top to the bottom wall. Velocity and temperature fluctuations are highly correlated, as expected at unit Prandtl number, which qualitatively confirms the Reynolds analogy between momentum and heat transfer. However, even at $\Pran=1$, the thermal fluctuations show sharper gradients in the flow because of the absence of the smoothing effect of the pressure gradient in the temperature equation~\citep{Pirozzoli_Bernardini_Orlandi_2016}. We also note that the streamwise velocity fluctuations are lower than the temperature fluctuations inside the pores, which can again be attributed to the pressure gradient in the momentum equation. At $\Pran=0.5,2$, we still observe a significant correlation between the large-scale structures of the temperature and velocity fields in the channel core, although marked differences in the temperature gradients become visible. We also find much higher temperature fluctuations in the pores, resulting from a thinner conductive sublayer.

\begin{figure}
    \centering
    \includegraphics[width = \textwidth]{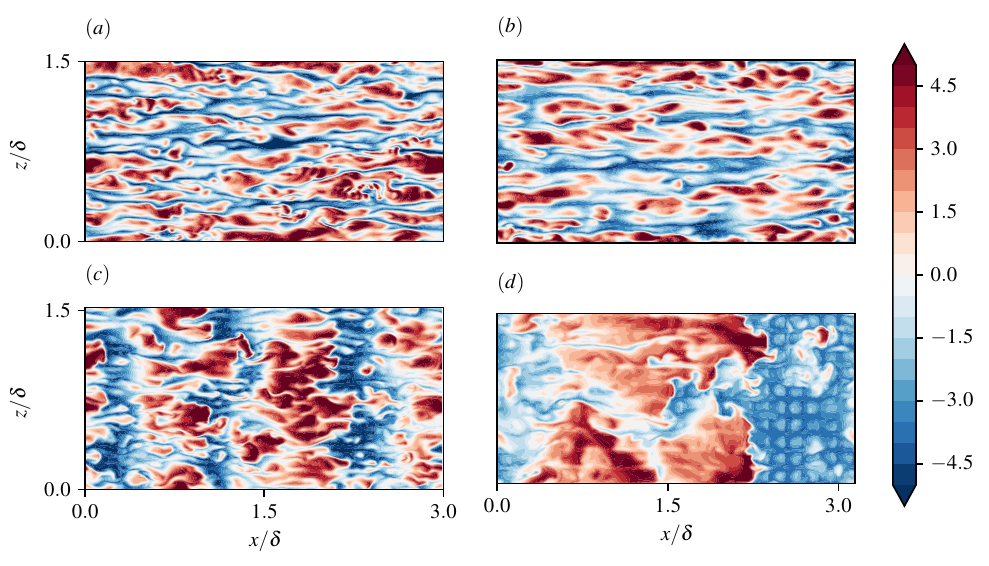}
    \caption{Instantaneous temperature fluctuations $\theta/\theta_{\tau}$ at $\Pran = 1$ in a wall-parallel plane at $y^+=15$ for $S700$ (\textit{a}), $L700$ (\textit{b}), $M700$(\textit{c}) and $H700$ (\textit{d}).}
    \label{fig:thetadash_xz}
\end{figure}

To further investigate how the thermal structures vary with changes in volume porosity, figure~\ref{fig:thetadash_xz} presents contours of temperature fluctuations at $Pr=1$ in a wall-parallel plane at $y^++\ell_T^+ \approx 15$ (where $\ell_T$ is the virtual origin of turbulence defined in section~\ref{virtualOrigin}). For the smooth-wall case S700, low- and high-speed streaky structures are clearly discernible, which is a characteristic signature of the near-wall cycle. 

With low wall porosity, for flow case $L700$, similar streaks are still visible, but they are shorter in the streamwise direction. \citet{breugem_boersma_uittenbogaard_2006} observed a similar shortening in the streamwise velocity variance with increasing permeability. This phenomenon is caused by the increase in wall-normal velocity fluctuations, also reported in the works of \citet{KUWATA2019} and \citet{Shahzad2022}. A major transition takes place when the porosity is increased to $71\%$, where short, blunt, blob-like structures start to appear, suggesting a departure from the near-wall cycle. At $87\%$ porosity, the thermal fluctuations highlight the appearance of spanwise-coherent structures that have been observed over porous surfaces~\citep{SUGA2011586, Manes_Poggi_Ridolfi_2011,Suga_Nakagawa_Kaneda_2017} and riblets~\citep{GARCÍA-MAYORAL_JIMÉNEZ_2011,endrikat_21}.

\subsection{Virtual origin} \label{virtualOrigin}

\begin{figure}
    \centering
    \includegraphics[width = \textwidth]{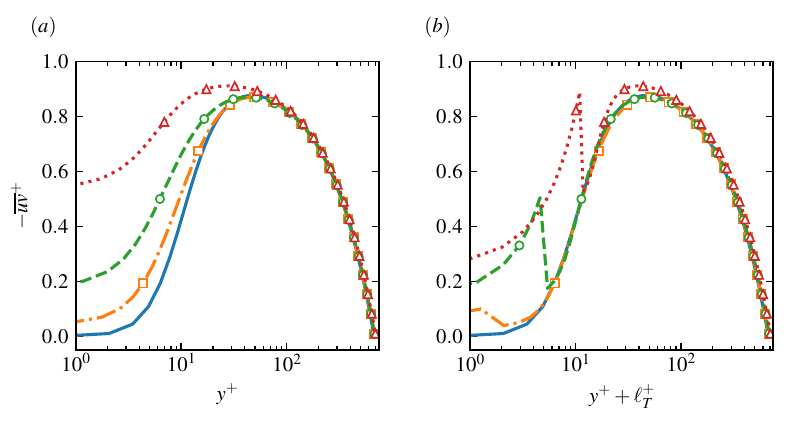}
    \caption{Reynolds shear stress $\overline{uv}^+$ as a function of the wall distance without (\textit{a}) and with (\textit{b}) the virtual-origin shift. Blue solid lines shows $S700$ case, orange dash-dotted lines shows $L700$, green dashed line shows $M700$ and red dotted line refers to $H700$ case.} 
    \label{fig:virtualOrigin700}
\end{figure}

Comparing smooth-wall statistics with those over a porous wall requires taking into account a proper wall-normal origin. Over a porous wall, the impermeability condition is relaxed; therefore, the location of the wall perceived by the flow is not at the interface between the free channel and the porous substrate, but is shifted into the porous medium. This shifted location, where the outer flow can be hypothesised to perceive the `wall', is known as the virtual origin. In the present work, we use the approach adopted by \citet{Ibrahim_Gómez-de-Segura_Chung_García-Mayoral_2021}, where we shift the Reynolds shear-stress profile by a virtual-origin shift $\ell_T$ to match the smooth-wall case at matched Reynolds number as closely as possible; see figure~\ref{fig:virtualOrigin700}. We note that, with this approach, we recover a good match between the porous- and smooth-wall cases, apart from the high-porosity cases, for which the Reynolds shear-stress distribution is inherently slightly different from the reference.

\begin{figure}
    \centering
    \includegraphics[width = \textwidth]{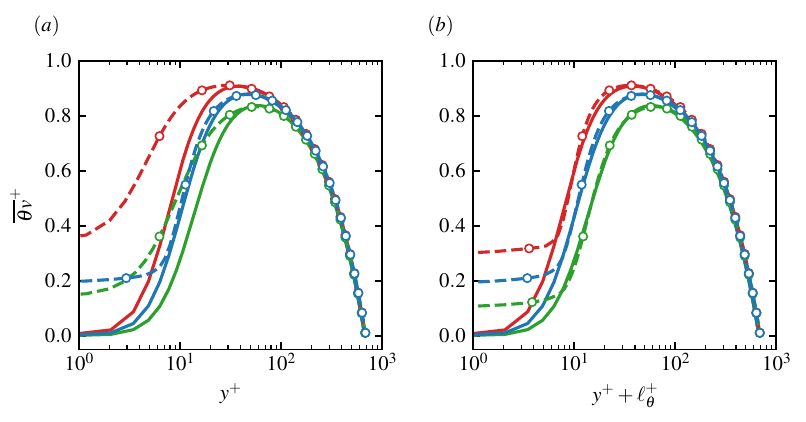}
    \caption{Turbulent heat flux $\overline{\theta v}^+$ profiles as a function of the wall distance without (\textit{a}) with (\textit{b}) the virtual origin shift Blue lines shows $\Pran=1$, green lines shows $\Pran=0.5$ and red line shows $\Pran=2$. Solid lines refer to $S700$ case and dashed lines refer to $M700$ case.} 
    \label{fig:virtualOrigin700_temp}
\end{figure}

Equivalently, the virtual origin for temperature is calculated by shifting the turbulent heat flux to match the smooth-wall case at matched Prandtl number, as shown in figure~\ref{fig:virtualOrigin700_temp}. At $Pr=1$, the virtual-origin shift for temperature coincides with that for momentum; however, for $Pr \neq 1$, this is not the case. The virtual-origin shifts for momentum and temperature for all the cases examined are summarised in table \ref{table:virtualorigin}, where differences due to the Prandtl number are evident.

\begin{table}
\centering
\begin{adjustbox}{max height=0.5\textheight, max width=1.5\textwidth}
\renewcommand{\arraystretch}{1.5} 
\begin{tabular}{ccccc}
\toprule
\textbf{Case} & $\ell_T^+$ & \multicolumn{3}{c}{$\ell_{\theta}^+$} \\
\cmidrule(lr){3-5}
 &  & $Pr = 2$ & $Pr = 1$ & $Pr = 0.5$ \\
\midrule
L260 & 1.3  & 1    & 0.6  & 1  \\
L500 & 2.0  & 0.5    & 0.7  & 1  \\
L700 & 3.0  & 2    & 2.0  & 2  \\
M260 & 2.7  & 3    & 2.0  & 3.2  \\
M500 & 4.7  & 5   & 4.7  & 5.4  \\
M700 & 5.1  & 5.75 & 5.6  & 6  \\
H260 & 10.2 & 8.25 & 10.2 & 10.3  \\
H500 & 12 & 8.5 & 12.0 & 12  \\
H700 & 12.5 & 9.5 & 12.1 & 14  \\
H1500 & 13  & 9.6   & 12.5   & 16   \\
\bottomrule
\end{tabular}
\end{adjustbox}
\caption{Virtual origin shifts for the mean velocity $\ell_T^+$ and for the temperature $\ell_{\theta}^+$ profiles.}
\label{table:virtualorigin}
\end{table}
\par

\subsection{Mean flow statistics}\label{Meanflow}

\begin{figure}
\centerline{\includegraphics{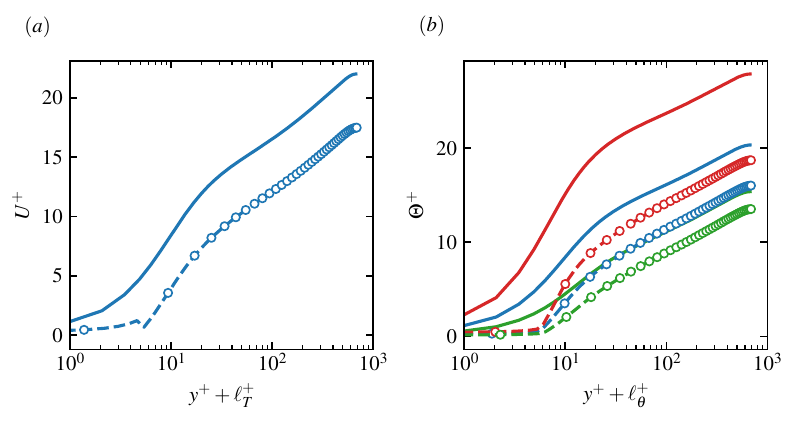}}
  \caption{(\textit{a}) Mean streamwise velocity and \textit{(b)} mean temperature profiles for case $S700$ (solid) and case $M700$ (dashed), for $\Pran=2$ (red), $Pr=1$ (blue) and $Pr=0.5$ (green).}
\label{fig:meanveltempM700}
\end{figure}

Figure \ref{fig:meanveltempM700} shows the mean streamwise velocity and mean temperature profiles for case M700 at different Prandtl numbers. The porous-wall velocity and temperature profiles exhibit positive downward shifts, $\Delta U^+$ and $\Delta\Theta^+$, relative to the smooth-wall data, indicating higher drag and heat transfer, both of which increase with volume porosity.

We analyse the friction and heat transfer coefficients at different Prandtl numbers to assess the performance of the porous layer in terms of turbulent mixing. For drag performance, we consider the friction coefficient $C_f=2\tau_w/(\rho u_b^2)$ and the friction factor $f=4C_f$, whereas for heat transfer performance we consider the Stanton number $\St=-q_w/\left( u_b \theta_m\right)$ and the Nusselt number $\Nu=\Rey_b\St\Pran$, where $u_b$ is the bulk flow velocity, $\theta_m$ is the mixed mean temperature, and $Re_b=2hu_b/\nu$ is the bulk Reynolds number. Even though these coefficients depend on the Reynolds number, they provide insight into the heat-transfer augmentation properties of the porous substrate.

\begin{figure}
    \centering
    \includegraphics{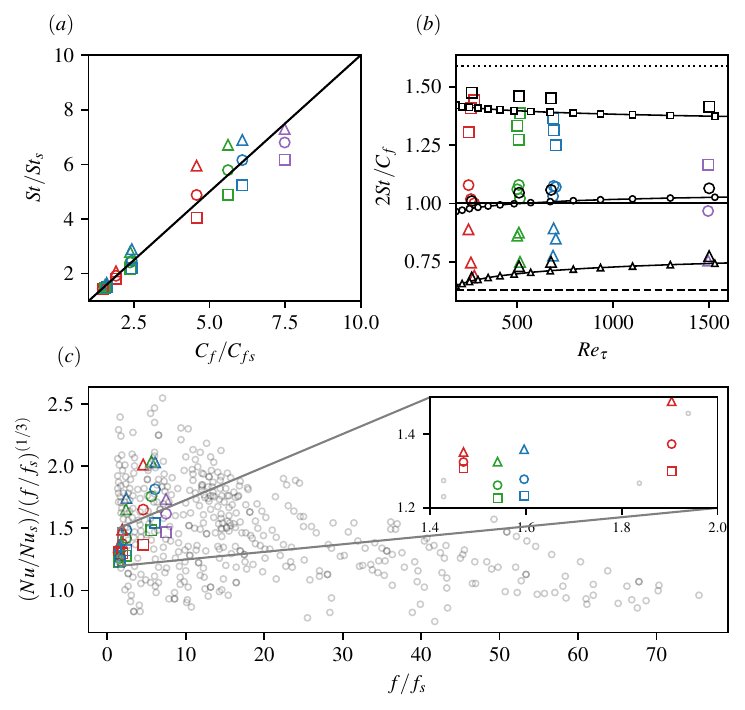}
    \caption{\textit{(a)} Bunker's plot,
     (\textit{b}) Reynolds analogy factor $2\St/C_f$ as a function of $\Rey_{\tau}$ for DNS data (symbols),
     Chilton-Colburn analogy ($2\St/C_f = \Pran^{-2/3}$, lines) and the Reynolds analogy from ThermoTurb (lines with symbols) at $\Pran=0.5$ (dashed), $\Pran=1$ (solid) and $\Pran=2$ (dotted).
     (\textit{c}) Performance enhancement factor as a function of the rough-to-smooth friction coefficient ratio, where gray symbols refer to roughness data from  \citet{Ligrani}.
    Symbols with black color refers to smooth wall data, whereas other colors refer to porous data at different Reynolds numbers: $\Rey_{\tau} = 260$ (red), $\Rey_{\tau}=500$ (green), $\Rey_{\tau}=700$ (blue), $\Rey_{\tau}=1500$ (purple). Different symbols refer to different Prandtl numbers $\Pran=0.5$ (squares),
    $\Pran=1$ (circles) and $\Pran=2$ (triangles).}
    \label{fig:Bunkerplot}
\end{figure}

We analyse these coefficients using different performance indicators, reported in figure~\ref{fig:Bunkerplot}. For smooth walls, figure~\ref{fig:Bunkerplot}\textit{(a)} shows the Reynolds analogy in the form of a Bunker plot. We find a non-monotonic trend with Reynolds number at $\Pran=1$ and $\Pran=2$, as the heat transfer first increases more than drag with increasing $\Rey_\tau$, and then decreases again at $\Rey_\tau=1500$. This behaviour is consistent with the current understanding of rough-wall heat transfer, and it will be discussed further later on. We further examine the relative enhancement of heat transfer with respect to drag using the Chilton--Colburn analogy \citep{ChiltonColburn1934}, which for a smooth wall reads $2St\,Pr^{2/3} = C_f$.

Figure~\ref{fig:Bunkerplot}\textit{(b)} shows the Reynolds analogy factor as a function of $\Rey_{\tau}$ for all flow cases, compared to the smooth-wall reference and to modelled smooth-wall data obtained from ThermoTurb~\citep[\href{www.thermoturb.com}{www.thermoturb.com}]{PIROZZOLI2024109544}. The plot confirms the non-monotonic trend in heat transfer observed previously, and it also shows that the Chilton--Colburn analogy for a smooth wall is fairly inaccurate for Prandtl numbers different from unity.

The performance enhancement factor (PEC), defined as $(\Nu/\Nu_s)/(f/f_s)^{1/3}$, is often used to characterise heat-transfer performance over rough walls~\citep{Ligrani} when accounting for constant power input. The PEC is above unity for all our flow cases, see Figure~\ref{fig:Bunkerplot}\textit{(c)}, which means that there is a heat-transfer enhancement benefit at matched pumping power. For comparison, figure~\ref{fig:Bunkerplot}\textit{(c)} also reports PEC data from \citet{Ligrani} for different types of roughness, with only the data points corresponding to swirl chambers showing higher PEC values. However, it must be noted that the data from \citet{Ligrani} come from experiments over rough surfaces in which the roughness material was mostly aluminium. In the present DNS, we impose an isothermal boundary condition, which assumes the thermal conductivity of the porous substrate to be infinite. Hence, for a fairer comparison, conjugate heat-transfer simulations would be necessary.

\subsection{Mean velocity deficit} \label{lengthscale} 
We aim to develop a model for the friction and heat transfer coefficients over porous surfaces. To this end, we need a framework for the synthetic generation of the mean velocity and temperature profiles in~\eqref{eq:vel_R} and~\eqref{eq:temp_R}, whose integration across the wall layer allows us to determine these coefficients. Models for the smooth-wall velocity~\citep{nagib_08} and temperature profile~\citep{PIROZZOLI2024109544} are available in the literature, whereas the main challenge remains finding suitable approximations for the velocity and temperature shifts $\Delta U^+(\ell^+)$ and $\Delta\Theta^+(\ell^+,\Pran)$.

Identifying a suitable length scale $\ell$ is the first step. This length scale should satisfy two properties, namely being physically relevant to the problem and leading to velocity and temperature shifts that are single-valued functions, possibly monotonic. For instance, the pore diameter is not a suitable length scale, as surfaces with the same pore diameter and different porosities result in different $\Delta U^+$~\citep{Shahzad_Hickel_Modesti_2023}. The Darcy permeability has often been used in the literature; however, it can result in a non-monotonic $\Delta U^+$, which is more difficult to model, especially in the transitionally rough regime~\citep{Shahzad_Hickel_Modesti_2023}. For turbulent flow grazing perforated plates, \citet{Shahzad2022,Shahzad_Hickel_Modesti_2023} showed that a length scale based on the wall-normal Forchheimer permeability results in the closest match with Nikuradse's data ~\citep{nikuradse1933}. 

\begin{figure}
  \centerline{\includegraphics{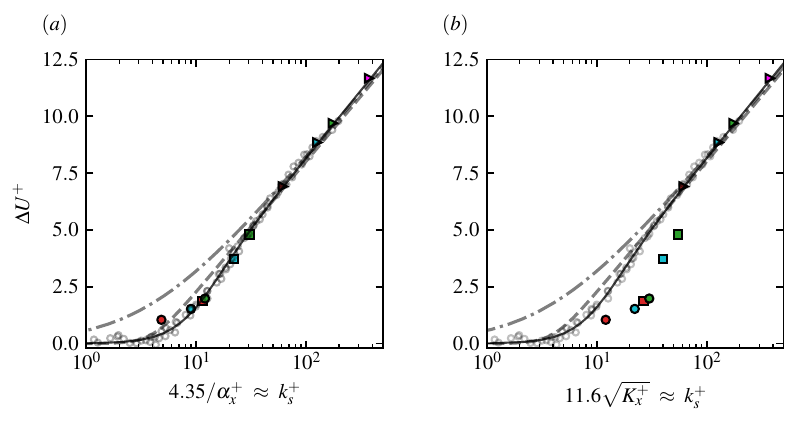}}
  \caption{Hama roughness function $\Delta U^+$ as a function of the streamwise viscous-scaled inverse of the (\textit{a}) Forchheimer permeability $4.35/\alpha_x^+ \approx k_s^+$, and (\textit{b}) Darcy permeability $11.6\sqrt{K_x^+} \approx k_s^+$. The dash-dotted line shows Colebrook's law~\citep{COLEBROOK1939}, the dashed black line shows the generalised roughness model of~\citet{Cheng_Pullin_Samtaney_2020}, the solid black line shows the model in equation~\eqref{eq:fit_deltau} and the black markers show Nikuradse's sandgrain roughness data~\citep{nikuradse1933}).}
  \label{fig:modellingdeltauandtheta}
\end{figure}

For the present study, the selection of a length scale is inherently more complex, owing to the non-zero permeability of the porous lattices in all three directions. Therefore, $\ell^+$ could be a combination of the Darcy and Forchheimer permeabilities in the $i$-th spatial direction,
\begin{equation}
    \ell^+ = A_i\sqrt{K_i^+} + B_i/\alpha_i^+.
    \label{eq:ks}
\end{equation}
To determine the coefficients in~\eqref{eq:ks}, we fit the present $\Delta U^+$ data to Nikuradse's sand-grain roughness data; thus, we effectively set $\ell^+=k_s^+$, where $k_s$ is the sand-grain roughness height. Having calculated the respective $k_s^+$ for the different cases, we utilize a regression fit with ~\eqref{eq:ks} to determine the dominant lengthscale. We use an analytical two-parameter fitting function as a model for Nikuradse's data, given by:
\begin{gather}
\Delta U^+(k_s^+) =
\frac{1}{p\,\kappa}\,\ln\!\left(1 + x(k_s^+)\right)
+ \left(\frac{1}{\kappa}\ln C + A - B\right)
\frac{x(k_s^+)}{1 + x(k_s^+)},
\label{eq:fit_deltau}
\end{gather}
where $x(k_s^+) = (k_s^+/C)^p$, with $p=2.28$ and $C=12.37$ as fitting constants, $A=4.17$ is the smooth-wall intercept, $B=7.9$ is Nikuradse's constant, and $\kappa=0.387$. This model is a modification of the one-parameter model proposed by~\citet{Cheng_Pullin_Samtaney_2020}, with the addition of a logistic damping function,
  \begin{gather}
     \Delta U^+ = \frac{1}{p\kappa}\mathrm{ln}(1+\beta(k_s^+)^p),  
     \label{eq:fit_pullin}
 \end{gather}
 where $\beta = \mathrm{exp}(p\kappa(A-B))$, $p=4$ is a fitting constant, $B=8$, and $A=4.5$. Both models approach the fully rough asymptote for large values of $k_s^+$, and the model by~\citet{Cheng_Pullin_Samtaney_2020} coincides with Colebrook's law~\citep{COLEBROOK1939} for $p=1$. Figure~\ref{fig:modellingdeltauandtheta} shows that equation~\eqref{eq:fit_deltau} accurately models Nikuradse's data across all values of $k_s^+$, and it is more accurate than Colebrook's law and slightly more accurate than equation~\eqref{eq:fit_pullin} in the transitionally rough regime.
 
Following the regression fit, $1/\alpha_x^+$ emerges as the dominant term, followed by $\sqrt{K_x^+}$. Fitting using only the streamwise Forchheimer permeability, we find $k_s^+\approx 4.35/\alpha_x^+$, whereas using only the Darcy permeability, $k_s^+\approx 11.6\sqrt{K_x^+}$; see figure~\ref{fig:modellingdeltauandtheta}. The figure shows that using the streamwise Forchheimer permeability results in a closer fit to the sand-grain roughness data from the transitional to the fully rough regime, permitting a simple modelling framework for all regimes.

\subsection{Mean temperature deficit} \label{tempdeficit}
\begin{figure}
{\includegraphics{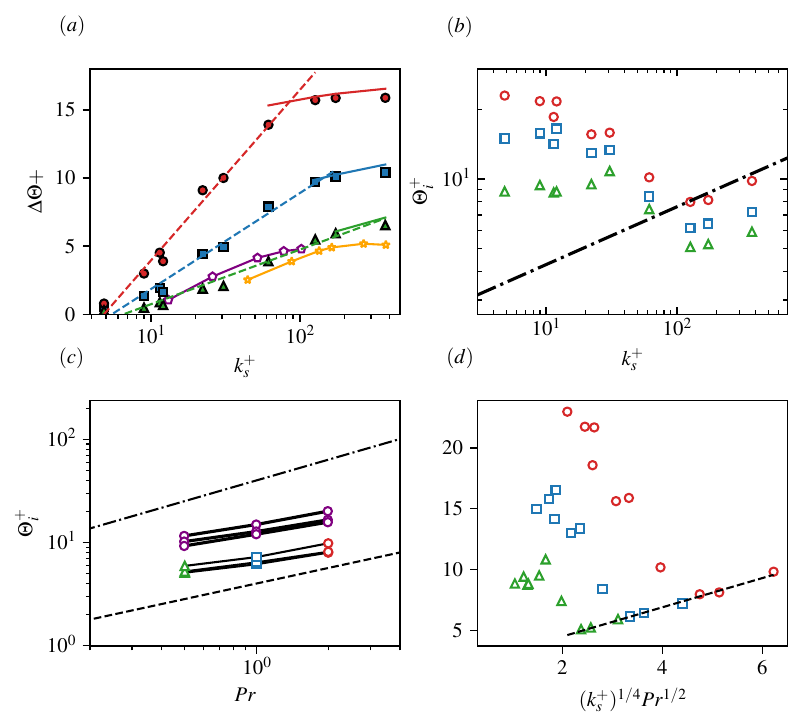}}
\caption{(\textit{a}) Temperature deficit $\Delta \Theta^+$ plotted as a function of $k_s^+\approx 4.35/\alpha_x^+$. The dashed line shows the empirical fitting~\eqref{eq:thetai_tr} for the transitional regime, and the solid lines show the model for the fully-rough regime. The purple markers show the data for irregular roughness at $\Pran=1$ \citep{PEETERS2019454} and the orange markers show the data for sinusoidal roughness at $\Pran=0.7$ \citep{MacDonald_Hutchins_Chung_2019}. (\textit{b}) Interfacial temperature $\Theta_i^+$ as a function of $k_s^+$, where the dash-dotted line shows $(k_s^+)^{1/4}$. \textit{(c)} Interfacial temperature ($\Theta_i^+$) as a function of the Prandtl number, where $\Pran^{0.5}$ (dashed) and $\Pran^{0.67}$ (dash-dotted). The purple markers shows the $\Theta_i^+$ data of sinusoidal roughness from \citet{Zhong_Hutchins_Chung_2023}. \textit{(d)} $\Theta_i^+$ plotted as a function of $(k_s^+)^{1/4} Pr^{1/2}$ compared to the fully rough asymptote~\eqref{eq:thetai_fr} (dashed line). Symbols refer to different Prandtl numbers: $\Pran=2$ (red), the $\Pran=1$ (blue) and $\Pran=0.5$ (green).}
\label{fig:modellingdeltatheta}
\end{figure}

Figure~\ref{fig:modellingdeltatheta}\textit{(a)} shows the temperature deficit $\Delta \Theta^+$ plotted against $k_s^+ = 4.35/\alpha_x^+$. For reference, DNS data by \citet{MacDonald_Hutchins_Chung_2019} for flow over 3D sinusoidal roughness at $\Pran=1$ are also reported, together with results from the DNS by~\citet{PEETERS2019454} over irregular roughness at $\Pran=0.7$. The surface roughness studied by~\citet{PEETERS2019454} closely resembles Nikuradse's data in terms of momentum deficit; however, the temperature deficit of the present cases does not match the data for irregular roughness, meaning that, in general, the factor $k_s\alpha_x$ differs for momentum and heat transfer.

The present data follow the same trend as the roughness data and seem to approach a plateau with increasing $1/\alpha_x^+$, which is more evident at $\Pran=2$, where we observe a slight decrease in $\Delta\Theta^+$ at the highest Reynolds number.

The problem of modelling the temperature deficit in the fully rough regime has often been cast in terms of the interfacial temperature $\Theta_i$ (see~\eqref{eq:Brutsaert}), which is related to the temperature deficit in~\eqref{eq:temp_R},
\begin{equation}
    \Theta_i = \frac{1}{\kappa_\theta}\log{y_i^+} + A_\Theta\left(\Pran\right) - \Delta\Theta^+.
\end{equation}

The main models for the interfacial temperature in the fully rough regime rely on a scaling of the form $\Theta_i\sim (k^+)^p\Pran^m$, where $m=0.5$ and $p=1/2$ in the model by~\citet{Owen_Thomson_1963}, and $p=1/4$ in the model by~\citet{Brutsaert1975}. The two models have been reviewed and verified by~\citet{Zhong_Hutchins_Chung_2023}, who reported that DNS data better follow the $p=1/4$ scaling. We compare this model with our DNS data in figure~\ref{fig:modellingdeltatheta}, where the interfacial temperature data are extracted at the point where the log law intersects the DNS temperature profile. The figure confirms the exponents given by~\citet{Brutsaert1975}, both for the Prandtl number and for the roughness height. In particular, figure~\ref{fig:modellingdeltatheta}(\textit{d}) shows an excellent match between the DNS data and the following fit in the fully rough regime,
\begin{equation}
\Theta_i^+ = 1.2 {k_s^+}^{1/4}\Pran^{1/2} + 2.1.
\label{eq:thetai_fr}
\end{equation}

 Using equation~\eqref{eq:thetai_fr} in \eqref{eq:Brutsaert}, we obtain a model for the temperature deficit, where we use $y_i\approx0.58k_s^+$, as measured from the DNS data. As also noted by \citet{Zhong_Hutchins_Chung_2023}, the scaling law \eqref{eq:thetai_fr} indicates that, at sufficiently high values of $k_s^+$, $\Delta \Theta^+$ will start to decrease and may even attain negative values, i.e. reduced heat transfer compared to smooth-channel simulations.

Regarding the transitionally rough regime, we do not expect a universal scaling for the interfacial temperature, as supported by the DNS data. Nonetheless, we attempt an empirical model for the temperature shift,
\begin{equation}
    \Delta \Theta^+ = C_1 (\Pran) \mathrm{log} (k_s^+) + C_2(\Pran),\quad 0 < k_s^+ < k_{s,fr}^+ ,
\label{eq:thetai_tr}
\end{equation}
where $C_1=2.46\Pran + 0.53$ and $C_2=-3.79\Pran - 1.53$ are fitting constants that depend on $\Pran$. $k_{s,fr}^+$ denotes the onset of the fully rough regime, for which we use $k_{s,fr}^+ \approx 100$ and $k_{s,fr}^+ \approx 70$ at $\Pran = 1$ and $\Pran = 2$, respectively. For $\Pran = 0.5$, there is no evidence of a fully rough regime.

\section{A model for friction and heat transfer coefficients} \label{lengthscale} 

Having introduced models for the velocity shift~\eqref{eq:fit_deltau} and for the temperature shift~\eqref{eq:Brutsaert} through equations~\eqref{eq:thetai_fr} and~\eqref{eq:thetai_tr}, we generate synthetic velocity and temperature profiles by shifting the smooth-wall reference. Here, we use the approach of~\citet{Pirozzoli_Modesti_2023}, in which the smooth-wall velocity and temperature profiles are modelled using their respective log laws in the inner region and a parabolic profile in the outer region. Hence, for the mean velocity profile we use
\begin{equation}
\begin{cases}
\begin{alignedat}{2}
 U^+  &= \displaystyle\frac{1}{\kappa}\log{\left(\eta\Rey_\tau\right)} + A -\Delta U^+, &\quad \eta < \eta^* \\ 
 U_e^+ - U^+ &= C \left(1-\eta\right)^2 + \Delta U^+, &\quad \eta > \eta^* 
\label{eq:velprof_model}
\end{alignedat}
\end{cases}
\end{equation}
and for the mean temperature profile
\begin{equation}
\begin{cases}
\begin{alignedat}{2}
\Theta^+ &=\displaystyle \frac{1}{\kappa_\theta}\log{\left(\eta\Rey_\tau\right)} + A_\theta\left(\Pran\right) - \Delta\Theta^+, &\quad \eta < \eta_\theta^*\\
\Theta_e^+-\Theta^+ &= C_\theta\left(1-\eta \right)^2+\Delta\Theta^+, &\quad \eta > \eta_\theta^*
\label{eq:tempprof_model}
\end{alignedat}
\end{cases}
\end{equation}
where we find $\eta^*,\eta_\theta^*$, $U_e^+$ and $\Theta_e^+$ by matching the functions and their first derivatives,
\begin{align}
\eta^* &=\displaystyle\frac{1}{2}\left(1-\sqrt{1-\frac{2}{C\kappa}}\right),\quad \eta_\theta^* =\displaystyle\frac{1}{2}\left(1-\sqrt{1-\frac{2}{C_\theta\kappa_\theta}}\right) \\ 
U_e^+ &= \frac{1}{\kappa}\log{\left(\eta^*\Rey_\tau\right)} + A - \Delta U^+ + C\left(1-\eta^*\right)^2\\
\Theta_e^+ &= \frac{1}{\kappa_\theta}\log{\left(\eta_\theta^*\Rey_\tau\right)} + A_\theta\left(\Pran\right) - \Delta\Theta^+ + C_\theta\left(1-\eta_\theta^*\right)^2,
\end{align}
and the constants calculated from the DNS data are $C=6.516$, $C_\theta=5.48$, $\kappa=0.387$, $\kappa_\theta=0.459$, and $A=4.8$.

\begin{figure}
\centerline{\includegraphics{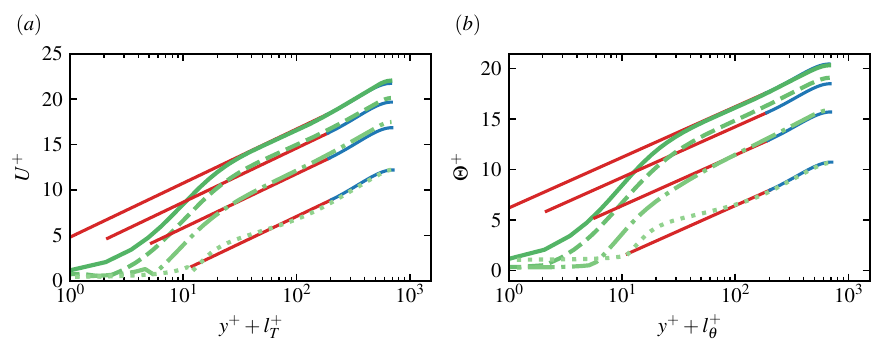}}
  \caption{Mean streamwise velocity (\textit{a}) and temperature (\textit{b}) profiles from DNS (dashed), compared to the composite profiles obtained matching the log law (red) with a wake (blue) as in~\eqref{eq:velprof_model} and ~\eqref{eq:tempprof_model} for cases $S700$ (solid), L700 (dashed), $M700$ (dash-dotted) and $H700$ (dotted).}
\label{fig:analyticalDNScompare}
\end{figure}

Figure~\ref{fig:analyticalDNScompare} shows the velocity and temperature profiles from DNS compared with the synthetic profiles in~\eqref{eq:velprof_model} and \eqref{eq:tempprof_model}, for values of $k_s^+=4.35/\alpha_x^+$ matching the current dataset. The modelled profiles match the DNS data with good accuracy, although minor discrepancies are observed for both velocity and temperature in the transitionally rough regime, where the models rely on empirical fittings.

The friction and heat transfer coefficients can be calculated by integrating the viscous-scaled temperature and velocity profiles,
\begin{align}
&C_f = \frac{2}{{u_b^+}^2},\quad \St=\frac{1}{u_b^+\theta_m^+},\\ 
&u_b^+ =\frac{1}{\Rey_\tau+\Phi h/\delta\Rey_\tau} \int_{0}^{\Rey_\tau}u^+\mathrm{d}y^+,\quad \quad \theta_m^+ =\frac{1}{u_b^+\left(\Rey_\tau+\Phi h/\delta\Rey_\tau\right)} \int_{0}^{\Rey_\tau}\theta^+u^+\mathrm{d}y^+.\label{eq:bulk}
\end{align}
In the model coefficients, we assume no flow below the porous medium and start integrating the profiles from $y^+=0$. The volume porosity in~\eqref{eq:bulk} appears because the bulk velocity and mean flow temperature are defined by integrating over the total fluid volume $V_f=L_xL_z(\delta + h\Phi)$.

We find the following expression for the friction coefficient,
\begin{equation}
    \begin{aligned}
        C_f = \frac{0.299 (1+\Phi h/\delta)^2}{\left(\log(\Rey_{\tau})-0.387 \Delta U^+ +0.932   \right)^2} .
    \end{aligned}
\label{eq:cf}
\end{equation}

Similarly, we obtain an analytical formula for the inverse of the Stanton number, 
\begin{equation}
\begin{aligned}
\frac{1}{St} = 
\frac{1}{(1+\Phi h/\delta)}&\Big[
5.63 \log^2(Re_{\tau})
+ \Delta U^+\big(-A_\theta\left(\Pran\right) - 2.18 \log(Re_{\tau}) + 2.27\big)\\
&
+ \Delta \Theta^+\big(-2.58 \log(Re_{\tau}) - 2.41\big)
+ \Delta U^+ \Delta \Theta^+ \\
&+ 2.58 A_\theta\left(\Pran\right)\ \log(Re_{\tau}) + 2.41 A_\theta\left(\Pran\right)
+ 0.03 \log(Re_{\tau}) + 1.30
\Big] .
\end{aligned}
\label{eq:stanton}
\end{equation}
We note that setting $\Phi=0$ in~\eqref{eq:cf} and~\eqref{eq:stanton} returns formulas for a generic rough wall, and for $\Delta \Theta^+=\Delta U^+=0$ we recover the smooth-wall formulas.

\begin{figure}
\centerline{\includegraphics{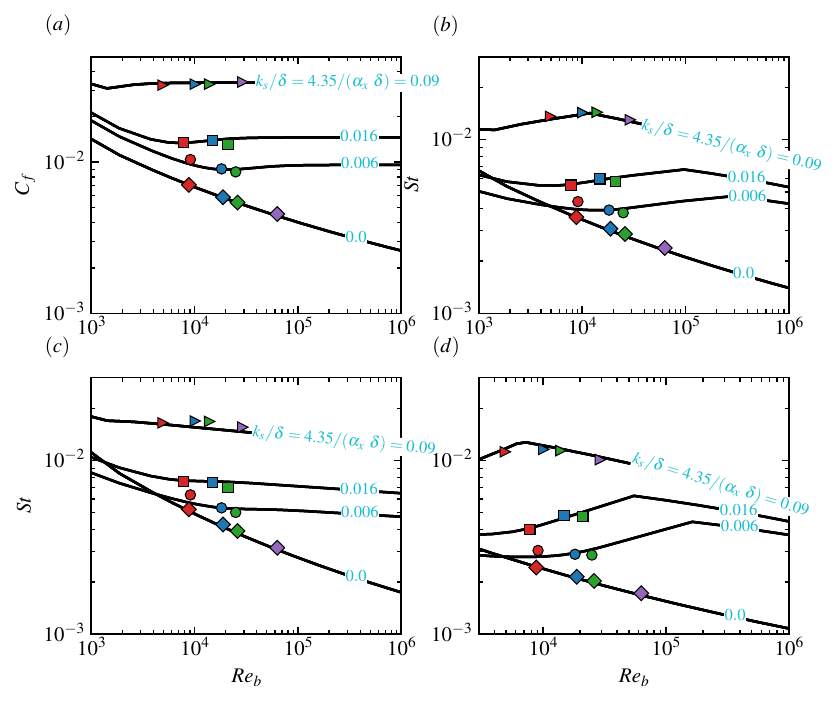}}
  \caption{Friction coefficient (\textit{a}) and Stanton number at $\Pran=1$ (\textit{b}),
  $\Pran=0.5$ (\textit{c}), and $\Pran=2$ (\textit{d}) as a function of the bulk Reynolds number for DNS data (symbols) and predictions (lines). Lines are evaluated at constant $k_s/ \delta = 4.35/(\alpha_x\ \delta)$. 
  Colors indicate different Reynolds numbers: $\Rey_\tau\approx260$ (red), $\Rey_\tau\approx500$ (blue)
  $\Rey_\tau\approx700$ (green) and $\Rey_\tau\approx1500$ (purple). Symbols indicate different porosity:
  smooth (diamond), low (circles), medium (squares) and high (triangles). The plots can be generated using the notebook provided \href{https://colab.research.google.com/drive/17ZMwTMYiabn7WCGGzigZqa17DPVXEAKG?usp=sharing}{here}.}
\label{fig:CfSt_Reb}
\end{figure}

\begin{figure}
    \centering
    \includegraphics{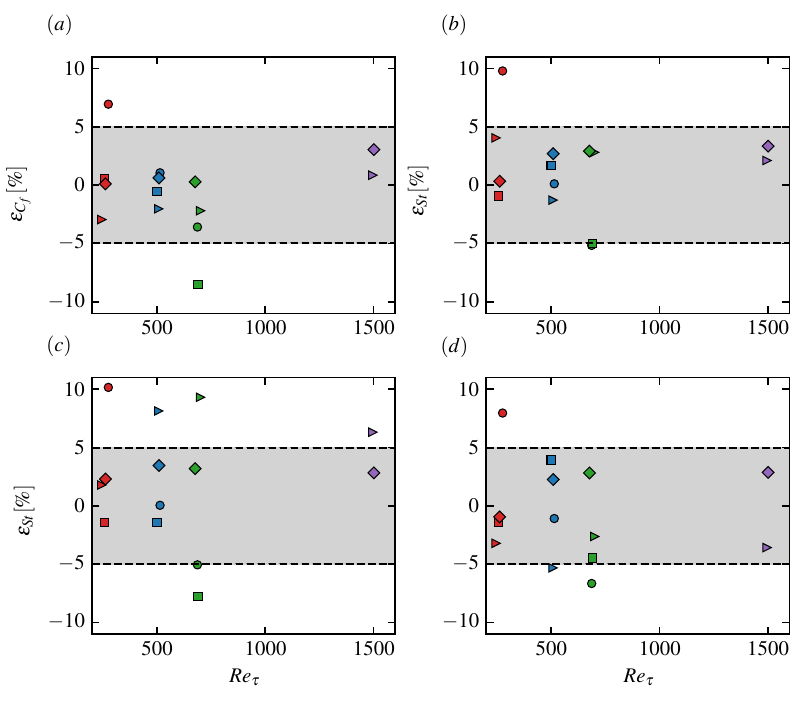}
  \caption{Percentage error on friction coefficient (\textit{a}) and Stanton number at $\Pran=1$ (\textit{b}),
  $\Pran=0.5$ (\textit{c}), and $\Pran=2$ (\textit{d}). The gray are represent the $\pm5\%$ error region. 
  Colors indicate different Reynolds numbers: $\Rey_\tau\approx260$ (red), $\Rey_\tau\approx500$ (blue)
  $\Rey_\tau\approx700$ (green) and $\Rey_\tau\approx1500$ (purple). Symbols indicate different porosity:
  smooth (diamond), low (circles), medium (squares) and high (triangles).}
    \label{fig:CfSterror}
\end{figure}

These predictive formulas for the friction coefficient and Stanton number are compared with the DNS data in figure~\ref{fig:CfSt_Reb}, showing excellent accuracy. The relative error is within about 5\% in the fully rough regime, see figure~\ref{fig:CfSterror}. These predictions are obtained using temperature and velocity shifts modelled with the Forchheimer permeability as the reference length scale. It is also possible to use the Darcy permeability, although this results in a much larger error in the transitionally rough regime, as shown in Appendix~\ref{ModellingKxplus}.

\section{Conclusions}\label{Conclusions}

We have studied turbulent forced convection in channel flows grazing over cubic-lattice porous substrates using a comprehensive set of direct numerical simulations (DNS) spanning Reynolds numbers \mbox{$\Rey_{\tau}=250$--$1500$}, Prandtl numbers $\Pran=0.5,1$, and $2$, and volumetric porosities $\Phi=50\%, 71\%$, and $87\%$. 

We focused in particular on the prediction of the friction and heat transfer coefficients by modelling the velocity and temperature shifts relative to the smooth-wall case. For the velocity shift, $\Delta U^+$, we find the emergence of a fully rough asymptote when either the streamwise Darcy permeability or the Forchheimer permeability is used as characteristic length scale. The Forchheimer scaling additionally yields very close agreement with Nikuradse's sand-grain data in the transitionally rough regime. The latter agreement is likely specific to the lattice geometry considered here, since no universal behaviour is expected in the transitionally rough regime. Nevertheless, the close correspondence with sand-grain roughness obtained using the Forchheimer scaling provides a particularly attractive framework for modelling the drag induced by cubic-lattice porous substrates. 

Considerably less is known about the temperature shift, $\Delta\Theta^+$, or equivalently the interfacial temperature, $\Theta_i$, in the fully rough regime, even for impermeable rough surfaces. The model of~\citet{Brutsaert1975} provides the most promising description available in the literature. The present DNS data show that this model can be extended to porous surfaces by replacing the roughness height with the streamwise Forchheimer permeability.

To our knowledge, this is the first time that a thermal fully rough regime has been reported and modelled for flows grazing over porous walls. According to our extension of the model of~\citet{Brutsaert1975}, the heat-transfer augmentation reaches a plateau and then decreases with $1/\alpha_x^+$, eventually becoming negative at very high Reynolds numbers. In real-world heat-transfer applications, however, cooling channels using gases as the working fluid (i.e. $\Pran\sim \mathcal{O}(1)$) operate at moderate friction Reynolds numbers, $\Rey_\tau\approx 1000$--$2000$, and therefore lie within the Reynolds-number range tested here. In this regime, simple cubic-lattice porous substrates appear to have a performance enhancement factor comparable to those of roughness patterns commonly used in heat-transfer enhancement applications, suggesting that porous substrates may provide a viable alternative.

We used the analytical velocity and temperature shifts to construct synthetic mean profiles. Integrating these profiles yields analytical formulas for the friction coefficient and Stanton number. We demonstrated that these predictions agree with the DNS data to within 5\% in the fully rough regime. The resulting framework provides a direct link between surface-induced shifts in the mean profiles and engineering quantities such as friction and heat transfer coefficients, and is readily applicable to wall-modelled simulations. This model framework is therefore not restricted to a particular porous substrate; the same formulas also apply to permeable or impermeable rough-wall cases, provided that a model for the velocity and temperature shifts is available.

\textbf{Declaration of interests.} The authors report no conflict of interest.

\vskip 1em
\textbf{Acknowledgements} The work by A.C. is supported by TU Delft AI labs and Talent Programme. A.C. acknowledges SURF (www.surf.nl) for the support in using the Dutch National Supercomputer Snellius through grants EINF-17240 and EINF-12445. D.M. acknowledges funding from the Italian Ministry of University and Research (MUR) through the FIS2 project HEATFORCE (grant no. FIS-2023-00346; CUP D53C25002050001).
\clearpage

\appendix
\section{ Immersed Boundary Method validation}\label{app:IBMvalidation}

\begin{figure}
\centerline{\includegraphics{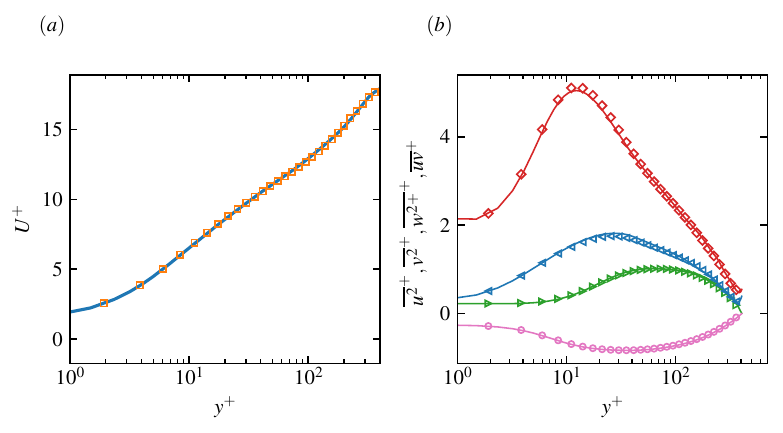}}
\caption{Comparison between present simulation (lines) DNS data of~\citet{MacDonald2018bars} (symbols): mean streamwise velocity (\textit{a}) and Reynolds stress components (\textit{b}), in viscous units.}
\label{IBMvalidation}
\end{figure}

We validate the immersed boundary method by reproducing the body-fitted DNS performed by~\citet{MacDonald2018bars}. The roughness consists of spanwise-aligned bars, and the computational box matches that of the reference study, $L_x \times L_z = 2.53\delta\times 0.77\delta$. The viscous-scaled spacing between the bars is $s^+ = 200$, the roughness height is $k^+ = 50$, and $\Rey_{\tau}=395$. We use a grid of $N_x \times N_y \times N_z = 800 \times 250 \times 80$ points. Figure \ref{IBMvalidation} shows an excellent match between the present simulations and the reference data from~\citet{MacDonald2018bars} for both the mean velocity and the Reynolds stresses.

\section{ Grid convergence study}\label{app:gridConvergence}
\begin{table}
\centering
\begin{tabular}{llllllllllll}
\hline
            Cases    & $L_x/\delta$   & $L_y/\delta$ & $L_z/\delta$    & $N_x$ & $N_y$ & $N_z$ & $\Delta x^+$ & $\Delta z^+$ & $N_\text{x}$ & $N_\text{z}$ & $N_\text{lig}$ \\
                \hline
Coarse  & 2.99208 & 2.6   & 1.527272 & 256   & 384   & 128   & 3.97         & 4.06         & 5             & 3             & 3              \\
Medium  & 2.99208 & 2.6   & 1.527272 & 512   & 384   & 256   & 1.98         & 2.03         & 10            & 6             & 6              \\
Fine    & 2.99208 & 2.6   & 1.527272 & 768   & 384   & 512   & 1.32         & 1.352        & 13            & 8             & 8         \\
\hline
\end{tabular}
\caption{Computational details of the grid convergence study: $L_x$, $L_y$ and $L_z$ represent the dimensions of the channel in streamwise, wall-normal and spanwise directions. $N_x$, $N_y$ and $N_z$ are the number of points in streamwise, wall-normal and spanwise directions. $\Delta x^+$ and $\Delta z^+$ are the grid spacings in streamwise and spanwise direction in plus units. $n_x$ and $n_z$ are the number of points per pore in the streamwise and spanwise directions, respectively, and $N_{lig}$ are the number of points per ligament.  }
\label{grid_ind}
\end{table}

\begin{figure}
\centerline{\includegraphics{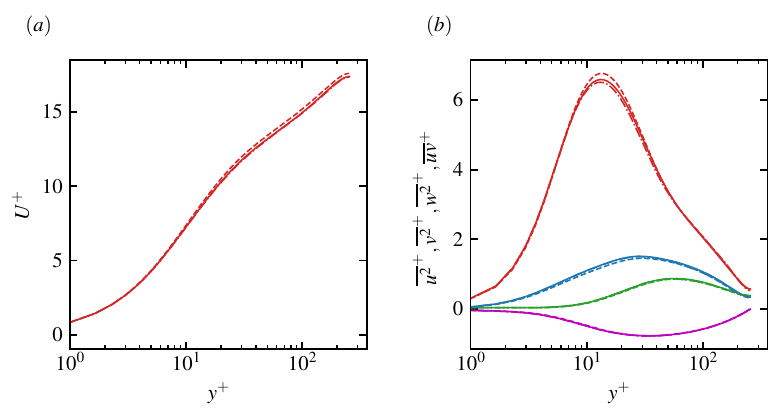}}
  \caption{Grid convergence study: Mean streamwise velocity (\textit{a}) and Reynolds stress components (\textit{b}) for coarse (dashed), medium (dash-dotted) and fine meshes (solid).}
\label{gridInd}
\end{figure}

We carried out a grid convergence study to determine the number of mesh points necessary to resolve the porous geometry, testing three meshes, as shown in table \ref{grid_ind}. A good agreement is visible between the medium and fine mesh, see figure \ref{gridInd}. Therefore, we use the medium grid, with 6-10 points per pore, to develop the DNS dataset.

\section{ Stokes flow simulations}\label{app:Openfoam}
OpenFOAM simulations are carried out to determine the permeability of the porous geometries. We carried out Stokes-flow simulations through porous unit cells at $Re_\text{p}$ (the pore Reynolds number) from 0.1 to 1000. An inlet--outlet geometry was set up with the cubic cell at the centre, with at least $50L$ distance between the inlet and outlet and the unit cell, where $L=0.3\delta$ is the dimension of the unit cell. Periodic boundary conditions are used on the sides of the unit cell. Further, no-slip boundary conditions are imposed on the cell itself. Inflow and outflow boundary conditions are imposed at the inlet and outlet, and pressure drops are calculated. A forward-Euler time scheme is employed using simpleFOAM, and the solution is allowed to converge until a steady-state residual of $10^{-7}$ is reached. Approximately $15$M mesh points are used after ensuring grid convergence. The Darcy--Forchheimer equation in the flow direction is used to calculate the permeabilities of the unit cells, given by:
\begin{align}
    \frac{\Delta P}{L} \frac{D^2}{\rho\nu U_t} = \frac{D^2}{K} + \sigma\alpha D Re_p,
\end{align}
where $\Delta P$ is the pressure drop, $D$ is the hydraulic diameter of the pores, $K$ is the linear permeability, $\alpha$ is the Forchheimer permeability, $\rho$ is the density, $\nu$ is the kinematic viscosity, and $U_t = \sigma U_p$, where $\sigma$ is the ratio of pore area to inlet area and $U_p$ is the pore velocity.

\section{Friction and heat transfer prediction with Darcy permeability}\label{ModellingKxplus}

\begin{figure}
\centerline{\includegraphics{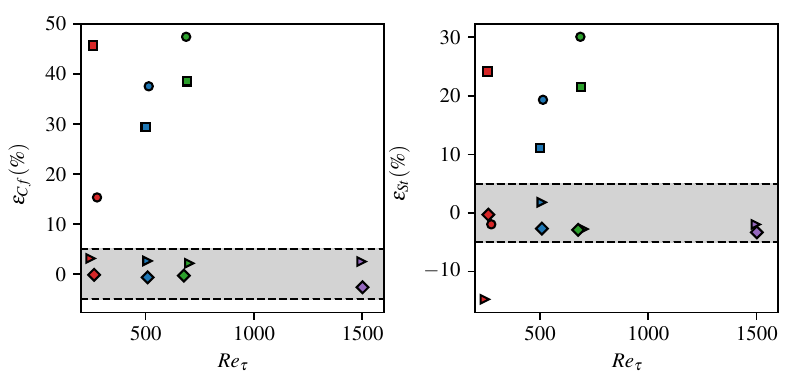}}
  \caption{Percentage error on friction coefficient (\textit{a}) and Stanton number at $\Pran=1$ (\textit{b}), obtained when using the streamwise Darcy permeability for predicting $\Delta U^+$ and $\Delta\Theta^+$. The gray are represent the $\pm5\%$ error region. 
  Colors indicate different Reynolds numbers: $\Rey_\tau\approx260$ (red), $\Rey_\tau\approx500$ (blue)
  $\Rey_\tau\approx700$ (green) and $\Rey_\tau\approx1500$ (purple). Symbols indicate different porosity:
  smooth (diamond), low (circles), medium (squares) and high (triangles).}
\label{fig:Kx_err}
\end{figure}

The temperature and velocity shifts can be modelled using either the Darcy or the Forchheimer permeability as reference length scales. In figure~\ref{fig:Kx_err}, we report the error in the prediction of the friction coefficient and the Stanton number when the streamwise Darcy permeability is used instead of the Forchheimer one. We find accuracy similar to that shown in figure~\ref{fig:CfSterror} in the fully rough regime, but considerably larger errors in the transitional regime.


\newpage


\newpage
\newpage

\bibliographystyle{jfm}
\bibliography{jfm}

\end{document}